\documentclass{ws06-procs}

\usepackage{graphicx}

\usepackage{amsfonts}

\begin{document}

\title{An X-ray investigation of Hickson 62}

\author{S. Sodani, E. De Filippis, M. Paolillo, G. Longo and M. Spavone}

\address{Dipartimento di Scienze Fisiche,Universit\`{a} Federico II,via Cinthia 6, I-80126 Napoli,Italy\\E-mail: silviasodani@tin.it}

\maketitle

\abstracts{We studied the X-ray properties of the Hickson Compact Group HCG~62, in
order to determine the properties and dynamic and evolutionary
state of its hot gaseous halo. Our analysis reveals
that the X-ray diffuse halo has an extremely complex morphological, thermal
and chemical structure. Two deep cavities, due to the presence of the AGN hosted by the central galaxy NGC 4778, are clearly visible in the group X-ray halo. The cavities appear to be surrounded by ridges of cool gas. The group shows a cool core associated with the dominant galaxy. In the outer regions the temperature structure is quite regular, while the metal abundance shows a more patchy distribution, with large ($\gtrsim 2$) Si/O and Si/Fe ratios.}

\section{Introduction}
Compact galaxy groups\cite{Hic97} are composed by a few ($>4$)
galaxies in close configuration, resulting in spatial
densities comparable to those found in the core of rich clusters: 
they hence represent a link between poor galaxy groups and rich clusters. 
Compact groups show small dynamical timescales (a
few $10^8$ yr), implying frequent interactions or complete merger
of the member galaxies, which are hardly observed in the majority
of them\cite{Hic97}. 
Such peculiar properties have cast doubts on the physical reality of these structures. 
Detailed spectroscopic and X-ray studies have
now brought a general agreement that these are real systems. The
implication is that a) significant evolution must have occurred in
most of these objects; b) to justify their observed number, the
dynamical timescales involved must be somehow diluted. One of the
favored scenarios suggests that CG are embedded in extended dark
matter halos which would significantly increase the dynamical
timescales. Such halos have been revealed in several groups
through extensive spectroscopic and X-ray surveys\cite{Zab98,Mul98}.

HCG~62 is one of the nearest CGs, with a redshift of $z=0.0137$
($\sim 58$ Mpc\footnote{$H_{0} = 70\ \mathrm{km\ s}^{-1}
\mathrm{Mpc}^{-1}$ ($\Omega_m = 0.3$, $\Omega_\Lambda= 0.7$}), and
it is composed of 4 early type galaxies. The group is dominated by
the two S0 galaxies NGC 4778 and NGC 4776 with a
projected separation of just 8 kpc. HCG~62 is embedded in a bright
X-ray halo extending out to $>300$ kpc with $L_X^{bol}\sim
10^{43}$ erg s$^{-1}$. 
The discovery of two large
cavities in its X-ray halo due to the presence of the AGN hosted by the central galaxy NGC 4778~\cite{Vrt00}, has recently drawn the attention
of the astronomical community on this nearby galaxy group.
Furthermore Spavone et al.(2006) have shown that this galaxy has a counter-rotating core and has thus experienced a recent accretion
event, indicating a possible link between the AGN activity
revealed by the {\em Chandra} X-ray observations and the merger
event.

\section{Morphological and Spectral Properties of the Diffuse Halo}
\begin{figure*}[t]
\centering
\resizebox{0.25\vsize}{!}{\includegraphics{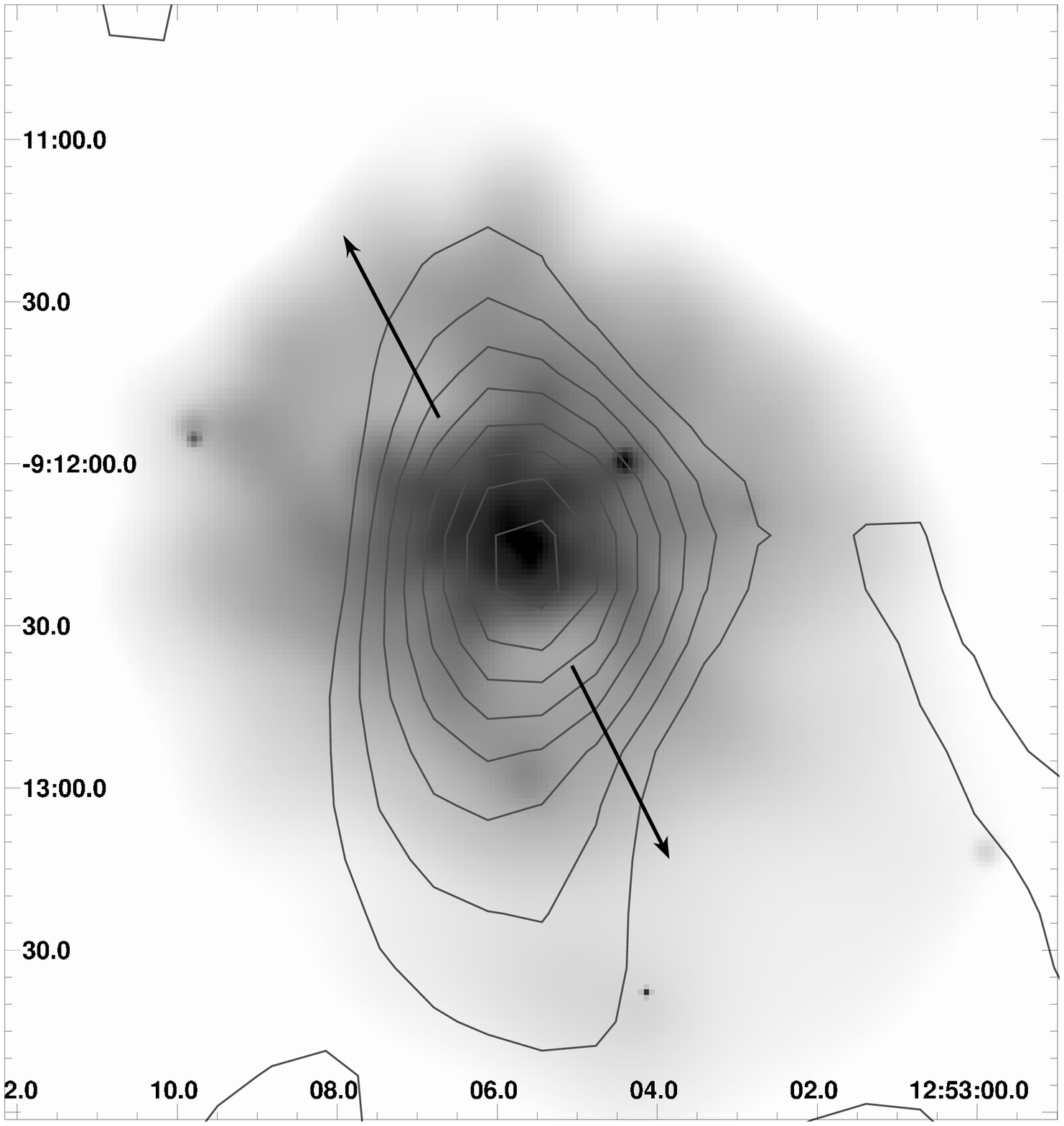}}
\resizebox{0.27\vsize}{!}{\includegraphics{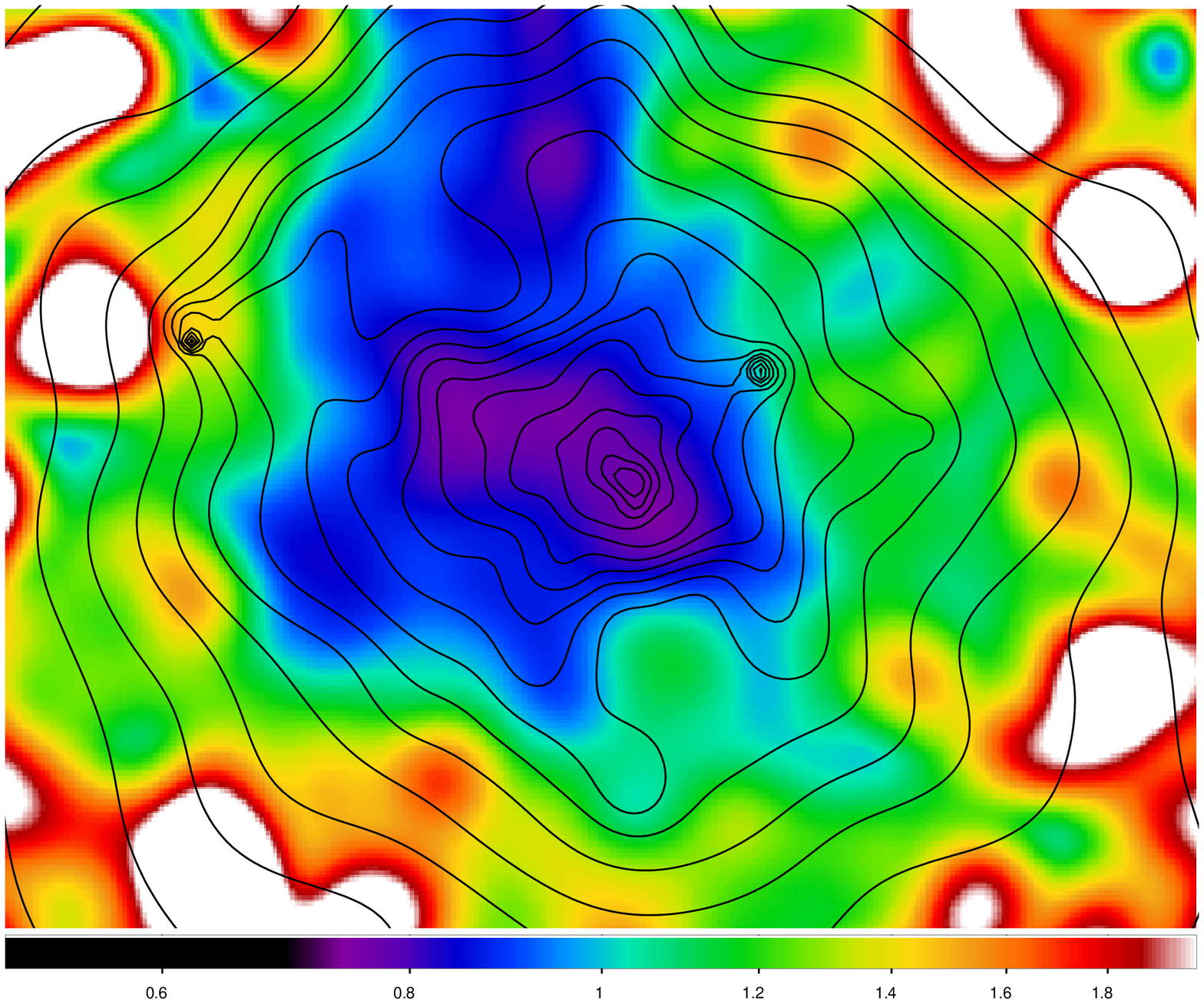}}
\caption{\textit{Left}:  Overlay of 1.4 GHz VLA radio contours onto the adaptively-smoothed X-ray image. Arrows indicate the orientation of the cavities. \textit{Right}: Overlay of the adaptively smoothed X-ray contours onto the Chandra temperature map. The temperature scale in keV is also shown.
}
\label{fig:img_opt.ps}
\end{figure*}

\begin{figure}[t]
\centering
\resizebox{0.27\vsize}{!}{\includegraphics{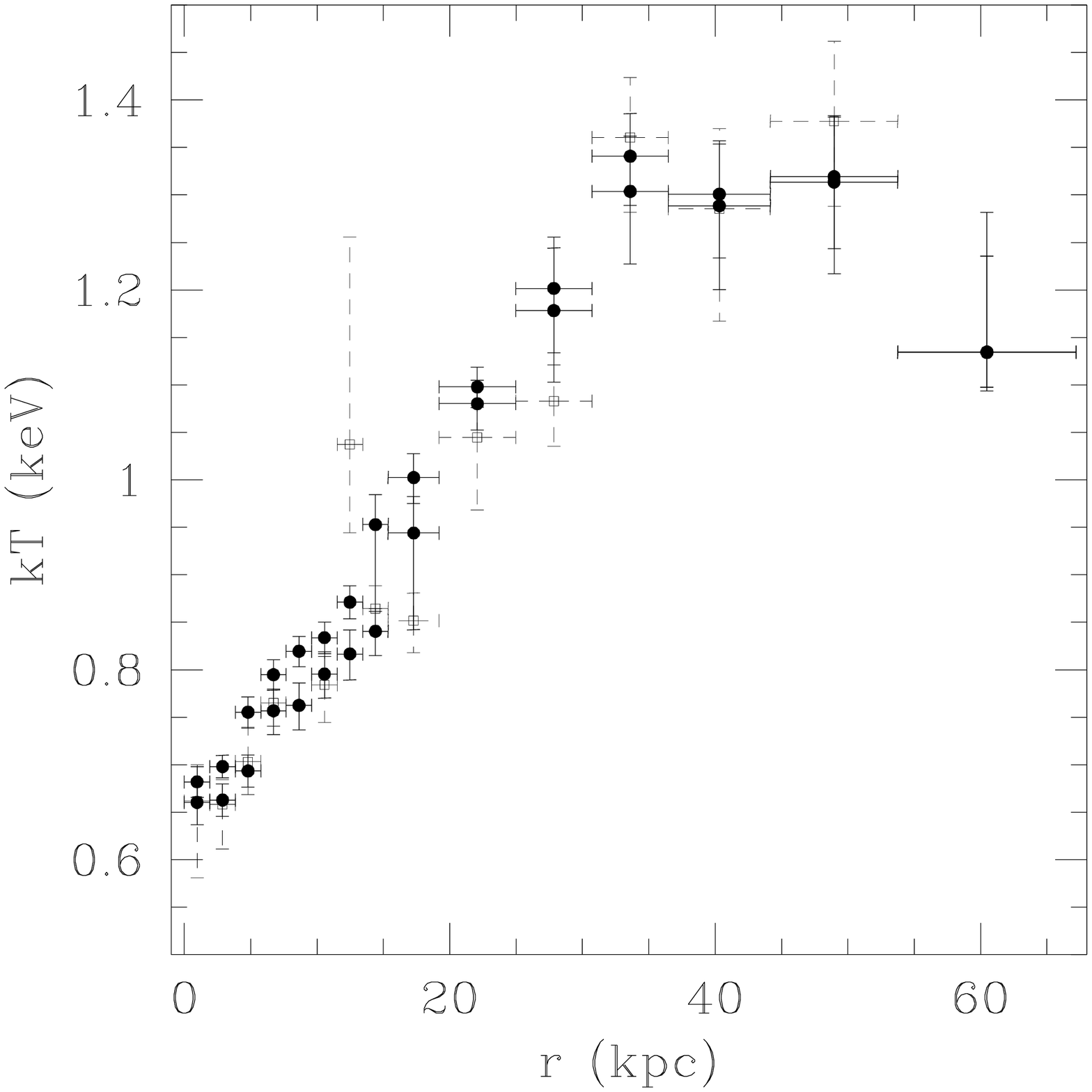}}\hspace{0.4cm}
\resizebox{0.27\vsize}{!}{\includegraphics{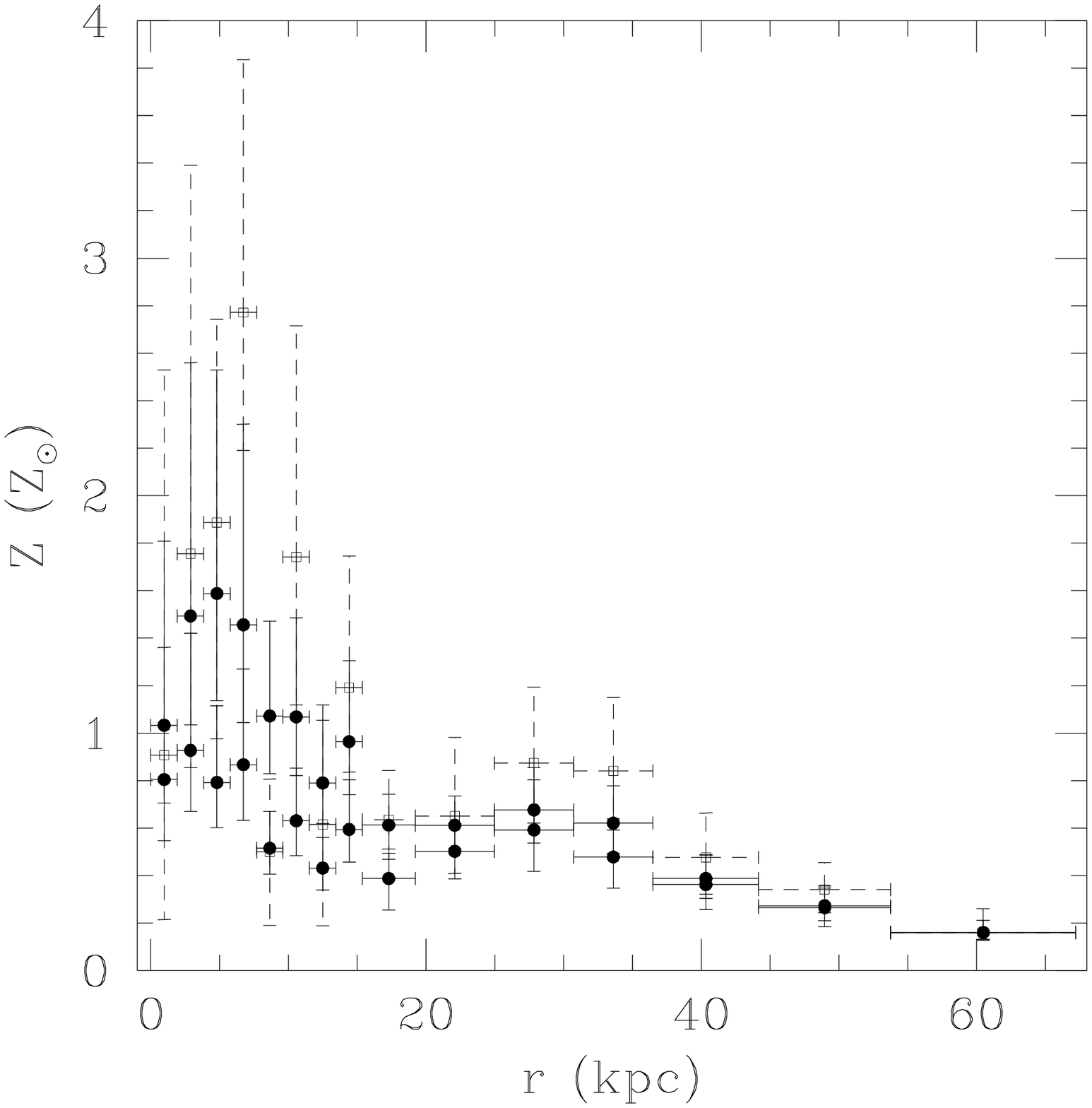}}\hspace{1cm}
\caption{Temperature and metal abundance profiles. Black (gray) solid circles with solid errorbars, represent the projected profile with thawed (frozen) HI. Empty squares and dashed lines show the deprojected profile.} \label{fig:spec_prof}
\end{figure}

In Fig.~\ref{fig:img_opt.ps} (left panel) we show 
the adaptively smoothed X-ray image from the ACIS-S3 chip.
The diffuse X-ray halo shows a NE-SW elongation and reveals the
presence of two deep cavities in the hot gas located
symmetrically with respect of the bright central galaxy
NGC 4778\cite{Vrt00}.
These cavities closely resemble those
discovered by {\em Chandra} in many galaxy clusters, 
produced by the relativistic radio-emitting jets produced by the AGNs hosted by the dominant
galaxy of the system~\cite{Birzan04,paolillo02,sch01}.\\
\noindent The simplest spectral model which could approximate, 
although with large residuals ($\chi^{2}_{\nu}\gtrsim 2$), the observed
X-ray emission within a radius of $3.6$ arcmin, 
is an absorbed plasma emission model
with two isothermal components. Both a MeKaL and an APEC model
in fact require the presence of a low-temperature $kT=0.76\pm0.02$ keV
component with supersolar (but unconstrained) metallicity and a
high-temperature component $kT=1.41\pm0.04$ keV with subsolar ($Z=
0.43\pm0.05$) metallicity. Leaving as free parameters Fe, O
and Si abundances while linking all other elements to Fe in their
solar ratios (using double VMeKaL
and VAPEC models), led to a measurable improvement in the model
fitting, implying unusually high Si/O and Si/Fe ratios ($\gtrsim 2$).
Results though show that both a single and a double temperature plasma
model, weighted over the whole group, are not adequate to describe
the properties of the diffuse emission, revealing a more
complex temperature and metallicity distribution.
Temperature and metal abundance profiles are shown in Fig.~\ref{fig:spec_prof}. The group temperature increases from kT$\sim 0.6\ \mathrm{keV}$ in the innermost regions to $\sim 1.3\ \mathrm{keV}$ at $\approx 30\ \mathrm{kpc}$. Past $60\ \mathrm{kpc}$ we observe a temperature decrease, measurable out to $\approx 200\ \mathrm{kpc}$ (as found from ROSAT data). The metal abundance profile reveals instead a metal rich group core, within $\mathrm{r}\approx 15\ \mathrm{kpc}$, but uncertainties, mainly due to the patchy abundance distribution in the central regions, are too large to be able to quantify the effect. The metallicity shows hints of a secondary peak at $\mathrm{r}\approx 30\ \mathrm{kpc}$, then slowly decreases going to larger radii.

A more accurate description of the physical properties of
the intragroup medium and of their spatial distributions has
required a two-dimensional spectral analysis of the diffuse X-ray halo.
In the right panel of Fig.\ref{fig:img_opt.ps} is shown the resulting temperature map, where different shades of gray reflect the different temperatures; superposed are the logarithmically spaced csmoothed X-ray contours. The map was obtained fixing the hydrogen column density to the Galactic value, the redshift to the nominal group value, and using a single MeKaL model in each cell. 
The group shows a cool core associated with the dominant galaxy,
in agreement with the temperature profiles discussed above. The cavities themselves appear to be surrounded by cooler gas, as already observed in several galaxy clusters~\cite{sch01}. In the outer regions the temperature structure is more regular, slowly increasing with the distance from the group center. The metal abundance shows instead a more patchy distribution.

\section{Gas Mass and Total Gravitating Mass Distributions}
\begin{figure}[t]
\centering
\resizebox{0.27\vsize}{!}{\includegraphics{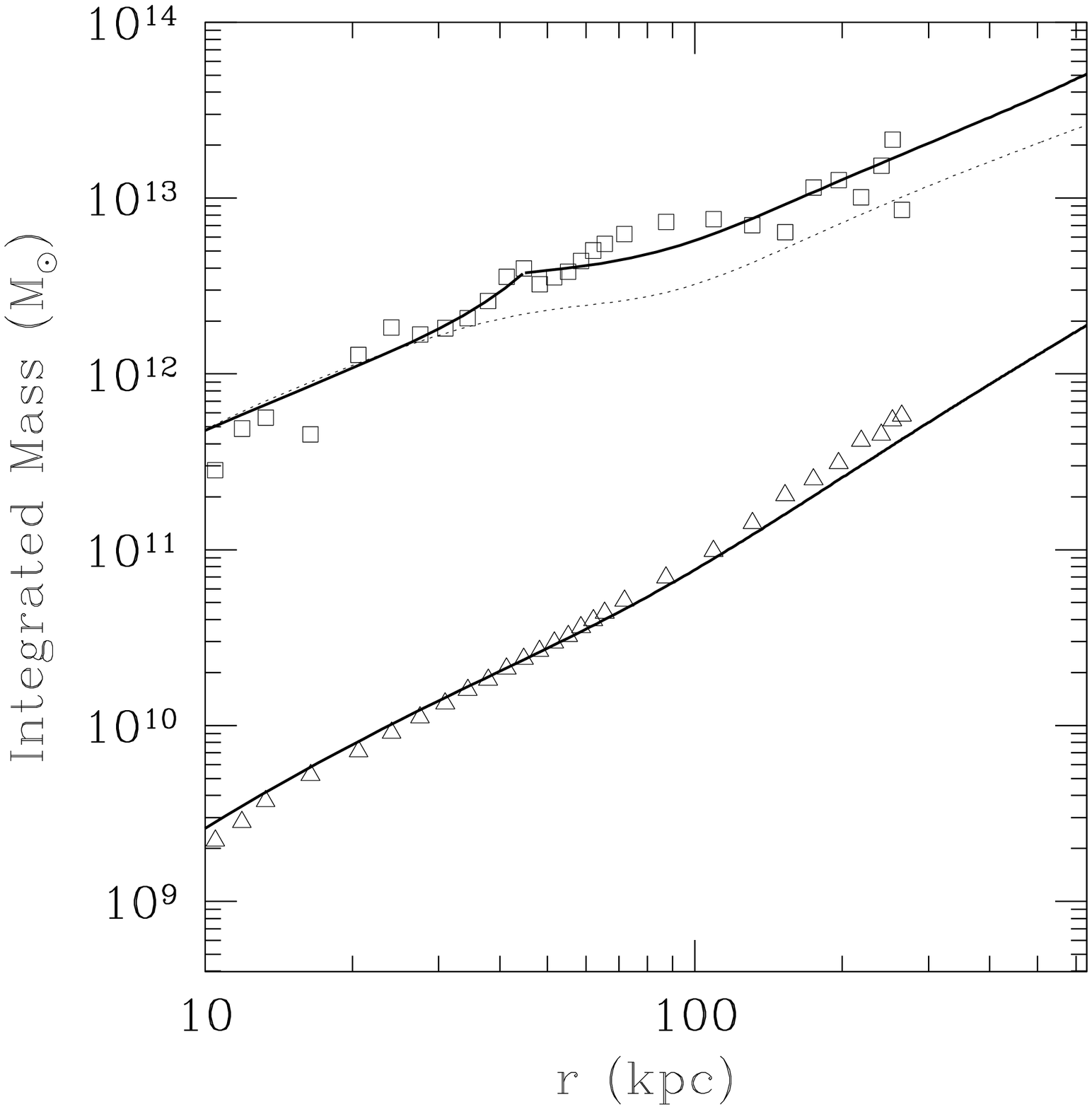}}\hspace{0.4cm}
\resizebox{0.27\vsize}{!}{\includegraphics{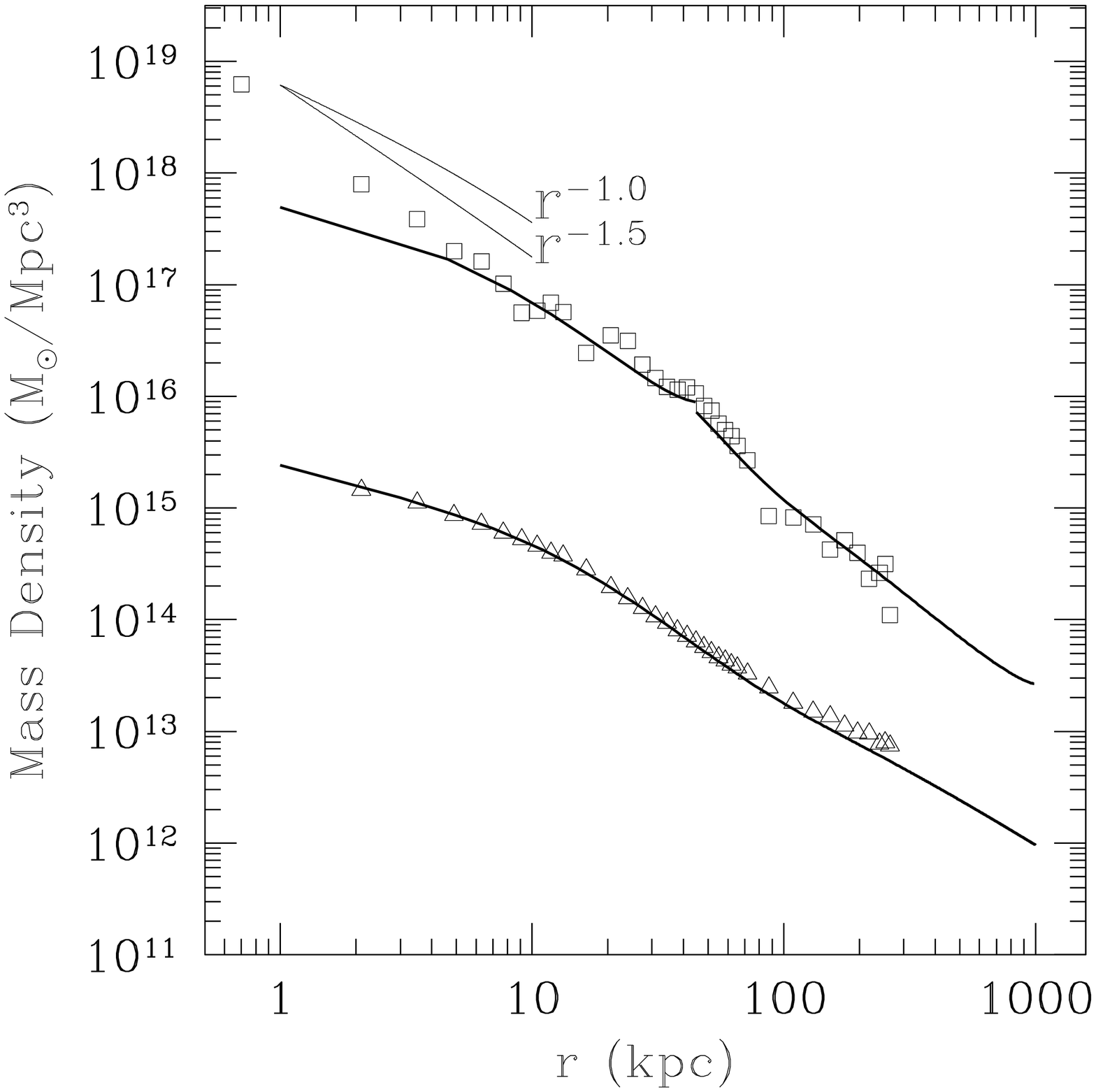}}
\caption{\textit{Left panel}: gaseous and total mass profiles (empty triangles and squares) derived from onion-peel deprojection. Solid and dotted lines represent the mass estimated through a double-$\beta$ model using the measured temperature profile or assuming an isothermal $kT=0.8\ \mathrm{keV}$ distribution. \textit{Right panel}: mass density profile and comparison with  NFW and Moore's models. The double-$\beta$ model (solid line) yields a fair fit, provided that the central $0.3$ arcmin are excluded, i.e. it has to be considered as a lower limit in the group innermost regions.} \label{fig:mass_prof}
\end{figure}

We derived the radial density distribution of the hot gas from
deprojection of the X-ray surface brightness profile,
assuming spherical symmetry and under the hypothesis of
hydrostatic equilibrium. We estimated the total gravitating
mass profile both through non-parametric deprojection or modeling
the surface brightness profile with a
double $\beta$-model.
The resulting
integrated gas mass and total gravitating mass profiles are shows
in Fig.~\ref{fig:mass_prof} ($\mathrm{M}_\mathrm{tot}(\mathrm{r}_{250\ \mathrm{kpc}})=8.5\times10^{12}\mathrm{M}_\odot$, $\mathrm{M}_\mathrm{gas}(\mathrm{r}_{250\ \mathrm{kpc}})=5.5\times10^{11}\mathrm{M}_\odot$). The plot highlights
that the assumption of isothermality leads to underestimate the
total group mass. In the right panel we compare the observed mass density profile with
two analytical models: the NFW and Moore's model, which have a central
density cusp proportional to r$^{-1}$ and r$^{-1.5}$, respectively. We found that the steeper Moore's profile provides a better fit to the data.

\section{Conclusions}
We analyzed the complex morphological, thermal and chemical structure of the HCG~62 X-ray halo. The halo shows a cool core with nearly solar abundances and hotter outskirts with subsolar metallicities. Our spectral analysis also suggest large ($\gtrsim 2$) Si/O and Si/Fe ratios. Interactions with the radio-emitting plasma has created two deep cavities in the X-ray diffuse halo; the cavities are surrounded by ridges of cool gas, as already noticed in richer galaxy structures. A steep Moore's profile provides a better fit to the inner regions of the total mass density profile.
Our results are supported by the independent analysis by Morita et al.(2006).

\end{document}